\documentstyle[12pt]{article}\textheight 220mm\textwidth 160mm
\begin{document}
\hspace*{11.6cm}KANAZAWA-93-2
\vspace*{0.7cm}
\begin{center}{\bf Gauge Equivalence in Two--Dimensional Gravity}
\end{center}
\vspace*{0.7cm}
\begin{center}{T. Fujiwara$\ ^{(1),(a)}$, Y. Igarashi$\ ^{(2),(b)}$,
J. Kubo$\ ^{(3),(c)}$, and T. Tabei$\ ^{(1)}$}
\end{center}
\vspace*{0.3cm}
\begin{center}
{\em $\ ^{(1)}$ Department of Physics, Ibaraki University,
Mito 310, Japan}\\
{\em $\ ^{(2)}$ Faculty of Education, Niigata
University, Niigata 950-21, Japan}\\
{\em $\ ^{(3)}$ College of Liberal Arts, Kanazawa University,
Kanazawa 920,
Japan }
\end{center}
\vspace*{0.3cm}
\begin{center}
ABSTRACT
\end{center}
Two-dimensional quantum gravity is identified as a
second-class system which we convert into a first-class
system via the
 Batalin-Fradkin (BF) procedure.
Using the extended phase space
method,
we then formulate the theory in most general class of
gauges.
The conformal gauge action suggested by David, Distler and Kawai
is derived from a first principle.
We find a local, light-cone gauge action
whose Becchi-Rouet-Stora-Tyutin invariance
implies Polyakov's curvature equation
$\partial_{-}R=\partial_{-}^{3}g_{++}=0$, revealing the origin
of the $SL(2,R)$ Kac-Moody symmetry.
The BF degree of freedom
turns out be dynamically active
as the Liouville mode in the conformal gauge, while
in the light-cone gauge the conformal degree of freedom
 plays that
r{\^ o}le.
The inclusion of the cosmological constant term in
both gauges and
  the harmonic gauge-fixing are also considered.

\vspace*{1cm}
\footnoterule
\vspace*{4mm}
\noindent
$^{(a)}$ E-mail address: tfjiwa@tansei.cc.u-tokyo.ac.jp  \\
$^{(b)}$ E-mail address: igarashi@ed.niigata-u.ac.jp\\
$^{(c)}$ E-mail address: jik@hep.s.kanazawa-u.ac.jp
\newpage
\pagestyle{plain}
\section{INTRODUCTION}
In spite of many efforts in studying two-dimensional (2D)
quantum gravity
[1--4], there are still some fundamental problems to be solved.
  First, the conformal gauge formulation developed by David,
Distler and
Kawai (DDK) \cite{d,dk} crucially
relies upon the validity of their conjecture on the functional
measure
that a Liouville action may be substituted for
the path-integral Jacobian
associated with the transition
to a translation invariant one.
There have been indeed several attempts \cite{fis,mmh} to justify the
conjecture, but a simple,
rigorous proof is lacking.
  Second, the origin of the $SL(2,R)$ Kac-Moody algebra
in the light-cone gauge, discovered by Knizhnik, Polyakov and
Zamolodchikov (KPZ)  \cite{pol1,kpz},
has remained obscure in the BRST quantization.
  Furthermore, there is no direct proof for the equivalence between
 the conformal gauge and light-cone gauge formulations,
although there are
various theoretical  indications for the equivalence \cite{mm}.

To overcome these problems, we formulate in this paper
Polyakov's string theory \cite{pol2} at sub-critical dimensions
\cite{ct} or 2D gravity
in most general class of gauges, according to our previous
proposal \cite{fikmt} to quantize the theory as an anomalous
gauge theory. \footnote{A similar program based on the anti-bracket
formalism \cite{anti} has been considered in Ref. \cite{gomis}.}
The strategy to achieve our aim is as follows:
We rely on the recent result \cite{fikm} on the
most general form of the BRST anomaly in 2D gravity
in the extended phase space (EPS) of Batalin, Fradkin and Vilkovisky
 (BFV) \cite{bfv}.
The anomaly found there expresses the fact that due to anomalous
Schwinger terms
the first-class (generalized) Virasoro
constraints have converted into second-class constraints.
This observation is our starting point to consider 2D gravity as an
anomalous gauge theory \cite{fs}.
We shall exactly follow the approach of Refs.\ \cite{mo,fik1}
to theories suffering from chiral gauge anomalies.
This method is based on the Hamiltonian formalism
developed by Batalin and Fradkin (BF)
\cite{bf} to quantize systems under second-class constraints.

The heart of the BF method \cite{bf} is to
rewrite a system under second-class constraints
into a gauge symmetric one by adding to the EPS the
compensating fields, the BF fields.
This idea can be simply extended \cite{mo,fik1}
to quantization of anomalous gauge theories \cite{fs},
because they can be regarded as second-class-constrained systems.
Remarkably, the re-conversion from the anomalous second-class
constraints back
into the effectively first-class ones can be
at least formally  performed without
considering gauge fixing.
One expects therefore that, once the corresponding
program has been successfully
applied to Polyakov's string,
one arrives at a quantum theory for 2D
gravity formulated in most general class of gauges.
This is indeed
possible as we will see.

A new and subtle feature in quantizing 2D gravity
as an anomalous gauge theory
 is that the compensating fields themselves contribute to the
anomaly
 in the gauge algebra. As a result, some of the
coefficients involved
in the anomaly-compensating
 mechanism
can not  be explicitly determined unless one fixes a gauge.
Therefore, the quantum nilpotency of the BRST charge should be
carefully examined for each gauge choice.
The desired gauge equivalence, formally ensured by the
 BFV theorem \cite{bfv,h}, can be achieved only in this manner.

We shall consider here three gauge choices: the
conformal \cite{d,dk},  light-cone \cite{pol1,kpz}
and harmonic gauge \cite{an} fixings.  The master action of BFV
contains two Liouville-type actions: One is
 for the Liouville mode arising from the compensating field,
and the other for the conformal factor
 of the metric, and they have a definite Weyl-transformation property.
  In the conformal gauge, the master action
becomes the effective action
suggested by DDK \cite{d,dk}, on one hand.
 In the light-cone gauge, on the other
hand, the master action becomes a
{\em local} action, a part of which looks like a Liouville action, and
the conformal mode acts
 as the gravitational analog of the
Liouville mode.
  The BRST invariance in this gauge-fixed theory naturally leads
to Polyakov's curvature equation \cite{pol1,kpz}
as well as to the $SL(2,R)$
Kac-Moody symmetry.
The KPZ condition can be obtained by investigating the
quantum nilpotency
 condition on the BRST charge.
So our local, light-cone gauge action contains the same information
as that of Polyakov's non-local action.
We also present a careful treatment of the
cosmological constant term in both the gauges.

In contrast to those gauge choices, the metric and ghost
variables become
 dynamically active in the harmonic gauge \cite{an}.
In the BFV formalism
this is equivalent
 to keep certain Legendre terms in the master action.

In section 2, we would like to briefly outline the
derivation \cite{fikm}
of the most general form of anomaly on the EPS in
Polyakov's theory. We apply in section 3 the canonical
method of Ref.\ \cite{fik1} to the present case: By an appropriately
chosen canonical pair of BF variables, we re-convert
the system into a first-class one.
The EPS variables are usually non-covariant. For covariant gauges,
e.g. for the conformal gauge, it is obviously convenient to use
covariant quantities. In section 4 we partially fix the
gauge to define
covariant quantities such as the metric variables and the covariant ghost
fields.
Section 5, 6 and 7 are respectively devoted to explicitly
consider the
gauge-fixed theory in the conformal \cite{d,dk},
light-cone \cite{pol1,kpz}, and  harmonic \cite{an} gauges.
Discussions
and summary are given in section 8.  In Appendix
we derive the
algebraic properties in the light-cone gauge
that are used in section 6.

\section{ANOMALY IN THE EPS}
Here we would like to briefly outline the derivation \cite{fikm} of
the most general form of anomaly on the EPS in
Polyakov's theory \cite{pol2}.
We begin by defining the EPS of the theory
described by the classical action
\footnote{We ignore the cosmological constant term here.
  It will be included later.}
\begin{eqnarray}
S_{\rm {X}} &=& -{1 \over 2} \int d^2 \sigma  \sqrt{-g}\,
g^{\alpha \beta}\, {\partial}_{\alpha} X^{\mu}\,
\partial_{\beta} X_{\mu}\ ,\\
\mbox{with}& &\alpha,\beta=0,1 ~\mbox{and}~ \mu=0,
\cdots,D-1\ ,\nonumber
\end{eqnarray}
where we mostly follow the notation of Ref.\ \cite{gsw}.
  We shall make an ADM decomposition for the 2D metric variables $
g_{\alpha\beta} $, parametrizing
\begin{eqnarray}
\lambda^{\pm} &= &
\frac{\sqrt{-g}\pm g_{01}}{g_{11}}\ , \quad  \xi ~=~ \ln g_{11}\ .
\end{eqnarray}
In this notation,
the Weyl transformation corresponds to a translation in the
$\xi$-variable.

The
conjugate momenta of $\lambda^{\pm}$ and $\xi$,
which we denote respectively
by $\pi^{\lambda}_{\pm}$ and  $\pi_{\xi}$,
vanish identically, yielding
the primary constraints
\begin{eqnarray}
\pi^{\lambda}_{\pm} \approx 0~ , \quad
\pi_{\xi} \approx 0~.
\end{eqnarray}
The Dirac algorithm further leads to the secondary constraints,
the Virasoro constraints
(we use the abbreviations $ \dot{f} = \partial_{\tau} f ~, ~
f^{\prime} = {\partial}_{\sigma}f~,~
\sigma^{\pm}=\tau\pm\sigma$, and $\partial_{\pm}
=\partial_{\tau}\pm\partial_{\sigma}$
with $\partial_{\pm}\,\sigma^{\pm}=2$.),
\begin{eqnarray}
\varphi_{\pm} &\equiv& {1 \over 4} ( P \pm X^{\prime})^2\ \approx 0\ ,
\end{eqnarray}
where $P_{\mu}$ denotes the conjugate momentum
of $X^{\mu}$ and given by
\begin{eqnarray}
P_{\mu} &=& -\sqrt{-g}g^{0\alpha}\ \partial_{\alpha}X_{\mu}\ ,
\end{eqnarray}
and the constraints $\varphi_{\pm}$ satisfy under
Poisson bracket the Virasoro algebra
\begin{eqnarray}
\{ \varphi_{\pm} ( \sigma )\ ,\ \varphi_{\pm}
( \sigma') \}_{\rm PB} &=& \pm
(\ \varphi_{\pm} ( \sigma )+ \varphi_{\pm} ( \sigma') \ )\
\delta^\prime ( \sigma - \sigma') \ ,\\
\{ \varphi_{+} ( \sigma )\ ,\ \varphi_{-} ( \sigma')
\}_{\rm PB} &=& 0 \nonumber\ .
\end{eqnarray}
So, at the classical
level, the theory is a system under the five first-class
constraints defined in Eqs. (3) and (4).

According to these five first-class constraints, we define
the EPS by adding
to the classical phase space the ghost-auxiliary field sector
which consists of the canonical pairs
\begin{eqnarray}
( {\cal C}^{\rm A}~ ,~ \overline{{\cal P}}_{\rm A} )~ ,~
( {\cal P}^{\rm A}~ ,~ \overline{{\cal C}}_
{\rm A} )~ ,~  ( N^{\rm A}\ ,\ {\cal B}_{\rm A} )~ ,
\end{eqnarray}
where  ${\rm A}\  ( = {\lambda}^{\pm},~\xi,~\pm)$
labels the first-class
constraints.
$ {\cal C}^{\rm A}$ and ${\cal P}^{\rm A}$ are the BFV ghost fields
carrying one unite of the ghost number,
$\mbox{gh}({\cal C}^{\rm A}) = \mbox{gh}({\cal P}^{\rm A}) =1$,
while $~~\mbox{gh}(\overline{{\cal P}}_{\rm A}) =
\mbox{gh}(\overline{{\cal C}}_ {\rm A}) = -1$
for their canonical momenta, $\overline{{\cal P}}_{\rm A}$ and
$\overline{{\cal C}}_ {\rm A}$.
The last canonical pairs in (7) are
auxiliary fields and carry no ghost number.
We assign $0$ to the canonical dimension of $
X^{\mu},~ {\lambda}^{\pm}$ and~ $\xi$,
and correspondingly $+1$ to $P_{\mu},~\pi^{\lambda}_{\pm}$, and
$\pi_{\xi}$. The canonical
dimensions of  ${\cal C}_{\lambda}^{\pm},~ {\cal C}^{\xi},~
\overline{{\cal P}}_{\pm}$,
$~\overline{{\cal P}}^{\lambda}_{\pm}$, and $\overline{{\cal P}}_{\xi}$
are fixed only relative to that of ${\cal C}^{\pm}$.
Let $ c~ \equiv \mbox{dim}({\cal C}^{\pm})$, we then have
\begin{eqnarray}
\mbox{dim}({\cal C}_{\lambda}^{\pm})~ =~
\mbox{dim}({\cal C}^{\xi})~=~1+c~ ,~
\mbox{dim}(\overline{{\cal P}}_{\pm})~ =~ 1-c~
,~  \mbox{dim}(\overline{{\cal P}}^{\lambda}_{\pm})~ =~
\mbox{dim}(\overline{{\cal P}}_{\xi})~=~-c~.
\end{eqnarray}

The BRST charge $Q$
 can be easily
constructed from the constraints given in Eqs. (3) and (4) as
\begin{eqnarray}
Q &=& \int d\sigma[~{\cal C}_{\lambda}^+ {\pi}^{\lambda}_{+}
+{\cal C}_{\lambda}^- {\pi}^{\lambda}_{-}
+{\cal C}^{\xi} \pi_{\xi}\nonumber\\
& &+{\cal C}^+(~\varphi_+
+\overline{{\cal P}}_{+}{\cal C}^{+\prime}~)
+{\cal C}^-(~\varphi_-
-\overline{{\cal P}}_-{\cal C}^{-\prime}~)
+{\cal P}^{\rm A} {\cal B}_{\rm A}~  ]\ ,
\end{eqnarray}
which
satisfies the super-Poisson
bracket (PB) relation
\begin{eqnarray}
\{Q~,~Q\}_{\rm PB} &=&0\ .
\end{eqnarray}
In quantum theory, the operator $Q$
should be suitably regularized, and
$Q^2$ expressed by the super-commutator $[Q~,~Q]/2$  may fail to
vanish due to an anomaly \cite{ko}.
In Ref.\ \cite{fik,fikm}, we imposed super-Jacobi
identities among super-commutation relations while
assuming that super-commutators can be expanded
in $\hbar$. Thus we assumed that the anomalous
commutator relation
\begin{eqnarray}
[Q\ ,\ Q] &= & i\hbar^2\Omega+{\rm O}(\hbar^3)\
\end{eqnarray}
 satisfies the super-Jacobi identity
\begin{eqnarray}
[Q\ ,[Q\ ,\ Q]\ ]&=&0\ .
\end{eqnarray}
In the lowest order in $\hbar$, Eq.\ (12) reads
\begin{eqnarray}
\delta\Omega &= &0\ ,
\end{eqnarray}
where $\delta$ is the BRST transformation given
by the Poisson bracket:
\begin{eqnarray}
\delta A &\equiv& - \{Q\ ,\ A\}_{\rm PB}\ .
\end{eqnarray}
Eq.\ (13) exhibits the consistency condition on $\Omega$ and
has to be solved to find
the most general  form of $Q^2$ in the
EPS.
The BRST non-trivial solution
to (13), which does not depend
on the choice of regularizations and gauges, is found to be \cite{fikm}
\begin{equation}
\Omega= K \int d\sigma\, [~{\cal C}^+\,
\partial_{\sigma}^{3}\, {\cal C}^+ -{\cal C}^-\,
\partial_{\sigma}^{3}\, {\cal C}^-~]
\end{equation}
for Polyakov's theory (1).
It is known from the explicit calculation
in the conformal gauge \cite{ko} that
\begin{eqnarray}
 K &=& \frac{(26-D)}{24\pi} \ ,
\end{eqnarray}
and that the $O(\hbar^3)$-term in Eq.\ (11) is absent.

Due to the very nature of our method to find anomalous Schwinger
terms,
the above result can be obtained without any reference
to a two-dimensional metric, and the result is
valid for any space-time
dimension, as emphasized in Ref.\ \cite{fikm}.
The $Q^2$-anomaly (11) expresses the anomalous conversion
of the first-class nature of the generalized Virasoro constraints
\begin{eqnarray}
\Phi_{\pm}&=&\{{\cal P}_{\pm}~,~Q\}_{\rm PB}~
=~ \varphi_{\pm} \pm 2 \overline{{\cal P}}_{\pm}
\partial_{\sigma} {\cal C}^{\pm} \pm \partial_{\sigma}
\overline {{\cal P}}_{\pm} {\cal C}^{\pm}\
\end{eqnarray}
into the second-class ones:
\begin{eqnarray}
[ \Phi_{\pm} ( \sigma ),~ \Phi_{\pm} ( \sigma') ] &=& \pm
i\hbar(\Phi_{\pm} ( \sigma ) + \Phi_{\pm} ( \sigma'))
\delta^\prime ( \sigma - \sigma')
\pm i\,K ~ \hbar^2 ~
\delta^{\prime\prime\prime}( \sigma - \sigma')\ .
\end{eqnarray}
The last terms in Eq.\ (18) are the anomalous Schwinger terms
in the Virasoro algebra.

The Hamiltonian plays a secondary role in the present case because the
canonical Hamiltonian $H_{\rm C}$ identically vanishes. Indeed,
if $H_{\rm C}=0$, the total
Hamiltonian in the BFV formalism is a BRST-commutator
\begin{eqnarray}
H_{\rm T} &=&\{Q~,~\Psi\}_{\rm PB}\ ,
\end{eqnarray}
where $\Psi$ is a gauge-fermion \cite{bfv}.

We fix the anomalous Schwinger term in
$[ Q\ ,\ H_{\rm T}]$  by
imposing the super-Jacobi identity \cite{fik,fikm}
\begin{eqnarray}
 2\ [Q\ ,\ [Q\ ,\ H_{\rm T}]\ ]+[H_{\rm T}\ ,\ [Q\ ,\ Q]\ ]&=&0\ .
\end{eqnarray}
Assuming that the Schwinger term again can be written as
\begin{eqnarray}
[ Q\ ,\ H_{\rm T}] &=& {i\over2}\hbar^2\Gamma+{\rm O}(\hbar^3)\ ,
\end{eqnarray}
one easily finds that the super-Jacobi identities among Poisson brackets
and the consistency condition on $\Omega$ given in (13) yield
\begin{eqnarray}
 \Gamma &=& \{\Omega\ ,\ \Psi\}_{\rm PB}\ .
\end{eqnarray}
For the standard form of the gauge fermion
\begin{eqnarray}
\Psi &=&\int d \sigma [\ \overline{{\cal C}}_{\rm A} {\chi}^{\rm A} +
\overline{{\cal P}}_{\rm A} N^{\rm A}\ ]\ ,
\ (A=\lambda^{\pm},~\xi,~\pm)\ ,
\end{eqnarray}
with the gauge-fixing functions ${\chi}^{\rm A} $,
we can compute the Poisson bracket on the right-hand side of Eq.\ (22)
unambiguously, because $\Omega$ contains only ${\cal C}^{\pm}$
(see Eq.\ (15)).
The only assumption we need is that $\chi$'s
do not depend on
$\overline{{\cal P}}_{\pm} $.
It leads to the
unique solution
\begin{eqnarray}
\Gamma &=& -2~K~\int d\sigma
[\ (\partial_{\sigma} N^+ \partial_{\sigma}^2 {\cal C}^{+}) -
(+ \rightarrow -)\ ]\ .
\end{eqnarray}
This result is independent of the gauge-fixing
functions ${\chi}$'s as long as the above assumption
 (which is about the weakest one imposed on $\Psi$)
is satisfied.
\section{SYMMETRIZATION}
Given the most general form of anomaly
in the EPS of the theory, expressed in Eqs.\ (15) and (24), we next
apply the BF algorithm  to re-convert the anomalous
system back into a gauge symmetric one, i.e.,
first-class under commutator
\footnote{From now on we suppress $\hbar$.}.
To this end, we introduce a canonical pair of
BF fields $(\theta, \pi_\theta)$ to cancel the anomaly,
and construct new effective Virasoro constraints
 ${\tilde{\Phi}}_{\pm}$ by adding an appropriate polynomial of
BF fields to $ \Phi_{\pm}$.  This new contribution is
fixed by requiring that  ${\tilde{\Phi}}_{\pm}$ satisfy the
anomaly-free Virasoro
algebra under commutator and reduce to $ \Phi_{\pm}$
(mod cobaoundary term) when the new fields are
set equal to zero.
The new constraints take the form \cite{ct,fik1}
\begin{eqnarray}
\tilde{\Phi}_{\pm}&\equiv& \Phi_{\pm} +~~
\frac{\kappa}{2}\, (~\frac{\Theta_\pm^2}{4}-\Theta_{\pm}^\prime~)
{}~+~~\frac{{\mu}^2}{2}~{\cal V}
 \ ,\\
\mbox{with}& &
\Theta_{\pm}=\theta^\prime \pm
\frac{2}{\kappa}~\pi_{\theta}~,\qquad {\cal V} =
\exp(\alpha~\theta)~, \nonumber
\end{eqnarray}
where $\Phi_{\pm}$ is given in Eq.\ (17), $\kappa$ and
$\alpha$ are coupling constants which will be determined later.
The last term
\footnote{For $\theta = 0$, the last term in Eq. (25)
reduces to a constant, which merely generates a coboundary term
in $\Omega$ of Eq. (15).
  Therefore, it does not affect the non-trivial solution (15).}
in Eq. (25)
would correspond to the inclusion of the
cosmological constant term, $-{\mu}^2\sqrt{-g}$,
in the string action (1).
The new BRST charge is given by
\begin{eqnarray}
\tilde{Q} &=& \int d\sigma[~{\cal C}_{\lambda}^+ {\pi}^{\lambda}_{+}
+{\cal C}_{\lambda}^- {\pi}^{\lambda}_{-}
+{\cal C}^{\xi} \pi_{\xi}\nonumber\\
& &+{\cal C}^+(~{\tilde{\varphi}}_+
+\overline{{\cal P}}_{+}{\cal C}^{+\prime}~)
+{\cal C}^-(~{\tilde{\varphi}}_-
-\overline{{\cal P}}_-{\cal C}^{-\prime}~)\nonumber\\
& &+{\cal P}^{\rm A} {\cal B}_{\rm A}~  ]\ ,
\ (A=\lambda^{\pm},\xi,\pm)\ ,
\end{eqnarray}
where
$$\tilde{\varphi}_{\pm}=\varphi_{\pm}+
\frac{\kappa}{2}\, (~\frac{1}{4}\Theta_\pm^2-\Theta_{\pm}^\prime~)
+ \frac{{\mu}^2}{2}~{\cal V}~.$$
 This $\tilde{Q}$ generates
the BRST transformations
\begin{eqnarray}
\delta X^{\mu} & =& {1 \over 2} [\ {\cal C}^{+}(P + X^{\prime})^{\mu}
+ {\cal C}^{-} (P - X^{\prime})^{\mu}\ ] \ ,\nonumber \\
\delta P^{\mu} & =& {1 \over 2} \partial_{\sigma}
[\ {\cal C}^{+}(P + X^{\prime})^{\mu}
- {\cal C}^{-}(P - X^{\prime})^{\mu}\ ]  \ ,\nonumber\\
\delta \theta &=&{1 \over 2} [\ {\cal C}^{+}
\,\Theta_{+}
- {\cal C}^{-} \,\Theta_{-}
+2{\cal C}^{+ {\prime}} -2{\cal C}^{- {\prime}}\ ] \ ,
\nonumber\\
\delta \pi_{\theta} &=&\frac{\kappa}{4}\partial_{\sigma}
[\ {\cal C}^{+}
\,\Theta_{+}
+ {\cal C}^{-}\,\Theta_{-}
+2{\cal C}^{+ {\prime}} +2{\cal C}^{- {\prime}}\ ]
 + \frac{{\mu}^2\alpha}{2}~({\cal C}^{+} +{\cal C}^{-})
{\cal V} \ ,\\
\delta {\cal C}^{\pm} & =& \pm {\cal C}^{\pm}
\partial_{\sigma} {\cal C}^{\pm}\ ,
\nonumber\\
\delta \overline{{\cal P}}_{\pm} & =&
- \tilde{\Phi}_{\pm}~,~
\delta {\lambda}^{\pm}~ =~ {\cal C}_{\lambda}^{\pm}~,
{}~\delta \xi~ =~
{\cal C}^{\xi} \ ,\nonumber\\ \delta{\pi}^{\lambda}_{\pm}
&=&\delta\pi_{\xi}\ =\  0\ ,\  \delta {\cal C}_{\lambda}^{\pm}\ =
\delta
{\cal C}^{\xi}\ =\ 0\ ,\nonumber\\
\delta\overline{{\cal P}}^{\lambda}_{\pm} &=& -\pi^{\lambda}_{\pm}\ ,\
\delta\overline{{\cal P}}_{\xi} = -\pi_{\xi}\ ,\  \delta N^{A}\ =\ {\cal
P}^{A}\ ,\ \delta\overline{C}_{A}\ =\ -{\cal B}_{A}\ ,\nonumber\\
\delta {\cal P}^{A}&=& \delta {\cal B}_{A}\ =\ 0~.\nonumber
\end{eqnarray}
The new Hamiltonian is defined by
\begin{eqnarray}
\tilde{H}_{\rm T}&=&\frac{1}{i}[\tilde{Q}~,~\Psi]\ ,
\end{eqnarray}
and
the new BRST charge $\tilde{Q}$ and
total Hamiltonian $\tilde{H}_{\rm T}$ are required to satisfy
the anomaly-free algebra
\begin{eqnarray}
[\tilde{Q}~,~\tilde{Q}]&=&[\tilde{Q}~,~\tilde{H}_{\rm T}]=0\ ,
\end{eqnarray}
for a suitable choice of $\kappa$ and $\alpha$,
which will be determined after having fixed a gauge.
\section{GAUGE-FIXED ACTION AND COVARIANTIZATION}
It should be emphasized
that
the BRST charge (26) is obtained prior to gauge-fixing.
The gauge-fixing appears in defining the total Hamiltonian
$~\tilde{H}_{\rm T}$
as in Eq.\ (28).
 The BRST invariant master action of
the theory (1) can then be written as \cite{bfv}
\begin{eqnarray}
S &=&\int d^2\sigma\, (~{\pi}^{\lambda}_{+}
\dot\lambda^++{\pi}^{\lambda}_{-}
\dot\lambda^-+\pi_\xi
\dot\xi+\pi_\theta\dot\theta+P_{\mu}\dot X^{\mu}\nonumber\\
& & +\overline{\cal P}_{\rm A} \dot{\cal C}^{\rm A}~)
-\int d\tau\,\tilde{H}_{\rm T}~.
\end{eqnarray}
(Except for the harmonic gauge discussed later,
we cancel the Legendre term
$\overline{\cal C}_{\rm A}\dot{\cal P}^{\rm A}
+{\cal B}_{\rm A} \dot{N}^{\rm A}$ in the action (30) by shifting
the gauge fermion as ${\Psi}\rightarrow{\Psi}+
\int d\sigma\, \overline{\cal C}_{\rm A} \dot N^{\rm A}$.)
 The BFV theorem \cite{bfv,h} formally
ensures that physical quantities in the quantum
theory, based on the action (30) along with $\tilde{Q}$ (26), are
$\Psi$-independent.

It is not straightforward to
recognize 2D gravity in the action (30) because
the geometrical meaning of the 2D metric variables is lost. To recover it,
we must go to configuration space.
This requires
elimination of various phase space variables by means
of the equations of motion, and for that we have to specify a gauge.
  In the standard form of the gauge fermion defined in Eq.\ (23),
we have five gauge conditions $\chi^{\rm A}\
({\rm A}=\lambda^{\pm},\xi,\pm)$.
To identify the 2D metric variables as well as the reparametrization
ghosts and the Weyl ghost, we use two of them to impose the
geometrization conditions \cite{fikm}
\begin{equation}
{\chi}_{\lambda}^{\pm}=\lambda^{\pm}-N^{\pm}~,
\end{equation}
while making  an (inessential) assumption that
$\chi^{\pm}$ and $\chi^{\xi}$ do not contain
$${\pi}^{\lambda}_{\pm},
\pi_{\xi}, \overline{{\cal C}}^{\lambda}_{\pm},
\overline{{\cal P}}_{\pm},
N_{\lambda}^{\pm},N^{\xi},
\overline{{\cal P}}^{\lambda}_{\pm}~,~
\mbox{and}~ {\cal B}^{\lambda}_{\pm}~.$$
One finds that
\begin{eqnarray}
\lambda^{\pm}&=&N^{\pm}~,
N_{\lambda}^{\pm} = ~ \dot\lambda^{\pm}~,~  N^{\xi} =
{}~ \dot\xi~,\nonumber\\
{\cal P}^{\pm}&=&{\cal C}_{\lambda}^{\pm} =~
\dot{\cal C}^{\pm}\pm{\cal C}^{\pm}\partial_{\sigma}N^{\pm}
\mp\partial_{\sigma}{\cal C}^{\pm}N^{\pm} ~,
\end{eqnarray}
can be still unambiguously derived. Then one can verify
that the covariant
variables defined by
\begin{eqnarray}
C^{0}&\equiv&{\cal C}^{0}/N^{0}~,~C^{1}={\cal C}^{1}-
N^{1}{\cal C}^{0}/N^{0}~,\\
C_W &\equiv&{\cal
C}^\xi-C^{0}N^{\xi}-C^{1}\partial_{\sigma}\xi
-2\,\partial_{\sigma}C^{0}N^{1}-2\,\partial_{\sigma}C~,\\
g_{\alpha\beta}&\equiv&
\left( \begin{array}{cc} -N^{+}N^{-} & (N^{+}
-N^{-})/2 \\
(N^{+}-N^{-})/2 & 1 \end{array} \right)\exp\xi~,
\end{eqnarray}
obey the covariant BRST transformation rules \cite{fuji}
\begin{eqnarray}
\delta g_{\alpha\beta}&=&C^{\gamma}\,
\partial_{\gamma} g_{\alpha\beta}
+\partial_\alpha C^{\gamma}\, g_{\gamma\beta}
+\partial_\beta C^{\gamma}\, g_{\alpha\gamma}+C_W\, g_{\alpha\beta}~,\\
\delta C^{\alpha} &=&C^{\gamma}\,
\partial_{\gamma} C^{\alpha}~,~
\delta C_{W} =C^{\gamma}\, \partial_{\gamma} C_{W}~,
\end{eqnarray}
where $ C^{\pm} = C^{0} \pm C^{1}$
and similarly for other quantities. Not that the $\tau $-derivatives in
Eqs.\ (36) and (37) are  exactly those which appear in
Eq.\ (32).

At this stage, we are left with three unspecified gauge conditions,
 which correspond to two reparametrization
and one Weyl symmetries.
  We shall consider in the following sections three gauges
to illustrate our formulation of 2D gravity,
which would clarify the relations to other approaches.

\section{CONFORMAL GAUGE}
The conformal gauge is defined by (31) and
\begin{equation}
\chi^{\pm}=N^{\pm}-\hat{N}^{\pm}~,
\end{equation}
where $\hat{N}^{\pm}$ and $\hat{\xi}$ are background fields which define
a background metric $\hat{g}_{\alpha\beta}$ (see Eq.\ (35)).
We substitute the gauge fermion $\Psi$ (23) with
the gauge-fixing functions (31) and (38) into the
master action (30) to obtain
the gauge-fixed action.
The momentum variables
can be eliminated by means of
 the equations of motions, e.g.,
\begin{eqnarray}
\pi_{\theta}&=&\frac{\kappa}{N^++N^-}[\,\dot\theta-
\frac{1}{2}(N^+ - N^-)\theta ' -(N^+ - N^-) '\,] \ ,\nonumber\\
P_{\mu}&=&\frac{2}{N^++N^-}[\,\dot{X}_{\mu}-
\frac{1}{2}(N^+ - N^-){X}_{\mu} '\,] \ ,\\
\overline{{\cal P}}_{\pm}&=&-\overline{{\cal C}}_{\pm}\ .\nonumber
\end{eqnarray}
Defining in terms of the BFV anti-ghosts, $\overline{\cal
C}_{\pm}$ and ${\overline{\cal C}}_{\xi}$, the covariant anti-ghosts
$\overline{b}_{\alpha \beta}$ (symmetric and traceless) by
\begin{eqnarray}
\overline{b}_{00}&=&  -(N^+)^2\,\overline{{\cal C}}_{+}-
(N^-)^2\,\overline{{\cal C}}_{-}\ ,\
\overline{b}_{01}= -N^+\,\overline{{\cal C}}_{+}+
N^-\,\overline{{\cal C}}_{-}\ ,\
\overline{b}_{11}=-\overline{{\cal C}}_{+}-\overline{{\cal C}}_{-}\ ,
\end{eqnarray}
and the Weyl anti-ghost $\overline C_W$
by
\begin{eqnarray}
\overline C_W &=& \frac{\overline{{\cal C}}_{\xi}}{\sqrt{-g}}\ ,
\end{eqnarray}
we obtain the conformal gauge
action in configuration space:
\begin{eqnarray}
S_{\rm CG}&=&S_{X}+S_{\phi}+S_{g}+S_{gf+gh}~, \nonumber\\
S_{\phi} &=& \int d^{2}\sigma\, \sqrt{-g}\,
[~-{\kappa\over2}\,({1\over2}\, g^{\alpha\beta}\,
\partial_\alpha \phi\,
\partial_\beta\phi+R\, \phi)~-{\mu}^2 V]~,\nonumber\\
S_{g}&=&\frac{\kappa}{2}
\int d^{2}\sigma\sqrt{-g}\, [~
{1\over2}\, g^{\alpha\beta}\,
\partial_\alpha\xi\, \partial_\beta\xi
-R\, \xi
 -2\frac{g_{11}}{g}\, \{({g_{01}\over
g_{11}} )^\prime\}^2~] ~,\nonumber\\
S_{gf+gh}&=&-\int d^{2}\sigma\, \{~{\cal B}_{\xi}\,
(N^{0} e^{\xi} - {\hat{N}}^{0}
e^{\hat{\xi}})
+{\cal B}_{+}\, (N^{+}-\hat{N}^{+})\nonumber\\
& &+{\cal B}_{-}\, (N^{-}-\hat{N}^{-})
+ \sqrt{-g}\,[~g^{\alpha\gamma}
\, \overline{b}_{\alpha\beta}\,
\nabla_{\gamma}C^{\beta}\nonumber\\
& &+\overline C_{W}\, (C_{W}+\nabla_{\alpha}C^{\alpha})~]~\}\ ,\\
\mbox{with}& &
\phi\equiv\theta-\xi,
\quad V\equiv \exp[\,\alpha\phi+(\alpha-1)\xi\,]~,\nonumber
\end{eqnarray}
where $\nabla_{\alpha}$ is the covariant
derivative, and $R$ the curvature scalar.

We next examine the quantum nilpotency of
 the BRST charge. To this end, we calculate the commutators of
the generalized Virasoro operators (25), where  all the operators
are  supposed to be normal ordered, and find that they
satisfy the algebra \cite{ct}
\begin{eqnarray}
[{\tilde{\Phi}}_{\pm}(\sigma)~,
{}~{\tilde{\Phi}}_{\pm}(\sigma^{\prime})
&=& \pm i\,({\tilde{\Phi}}_{\pm}(\sigma)
        +{\tilde{\Phi}}_{\pm}({\sigma}^{\prime}))
\delta^\prime
 ({\sigma}-{\sigma}^{\prime})
\nonumber\\
& & \pm i\, \frac{{\mu}^2}{2} \left(\alpha -
\frac{{\alpha}^2}{4\pi\kappa} -1
\right) ({\cal V}(\sigma) + {\cal V}({\sigma}^{\prime}))
\delta^\prime ({\sigma} - {\sigma}^{\prime})
\nonumber\\
& & \mp i \,\left( \frac{D+1-26}{24\pi} + \kappa \right)
 \delta^{\prime\prime\prime} (\sigma -{\sigma}^{\prime})~, \\
{}
[{\tilde{\Phi}}_{+}(\sigma)~,~{\tilde{\Phi}}_{-}({\sigma}^{\prime})~]
&=&
 i\, \frac{{\mu}^2}{2} \left(\alpha -
\frac{{\alpha}^2}{4\pi\kappa} -1 \right)
 {\partial}_{\sigma} {\cal V}(\sigma)~\delta
(\sigma -{\sigma}^{\prime})~.
\nonumber
\end{eqnarray}
Note that the BF
fields non-trivially contribute to the commutator anomaly by one
unit of the
 central charge in addition to the
$\kappa$-dependent classical contribution.
  The coupling relations needed for the closure of the algebra
and hence to ensure the nilpotency of the BRST charge are
\begin{eqnarray}
1 &=&\alpha - \frac{{\alpha}^2}{4\pi\kappa}~ ,\\
\kappa &=& \frac{(25-D)}{24\pi}~,
\end{eqnarray}
which lead to
\begin{eqnarray}
\alpha &=& {\alpha}_{\pm} = \frac{25-D \pm \sqrt{(25-D)(1-D)}}{12}~,
\end{eqnarray}
in accord with the result of Ref. \cite{dk}.

The conformal-gauge action given in Eq. (42) contains two
Liouville-type modes, $\phi$ and  $\xi$.
The BRST transformation of $\phi$ is covariant and given by
\begin{equation}
\delta \phi =  C^{\alpha}{\partial}_{\alpha} \phi - C_{W}~,
\end{equation}
and so it plays the r{\^ o}le of the conformal degree of freedom.
Note also that $V$ appearing in the cosmological constant term
 transforms as a world scalar
\begin{equation}
\delta V =  C^{\alpha}{\partial}_{\alpha} V - C_{W} V~.
\end{equation}
On the contrary to $\phi$ and $V$,
the $\xi\,(=\ln g_{11})$ is a non-covariant object,
 and
$S_{g}$ is a  non-covariant expression.
The origin of $S_{g}$ is related to
the fact that the manifest 2D covariance is violated
in the class of regularization schemes
\footnote{The normal-ordering prescription,
for example, is such a scheme.}
we approve.
In order to restore the 2D covariance,
one has to add an appropriately chosen non-covariant counterterm
to the action.
$S_{g}$ is nothing but this counteraction,
and  therefore, the present approach has
a built-in mechanism to keep the 2D covariance \footnote{In the
operator language,
$S_{X}+ S_{g}$ is thus reparametrization invariant, but not
Weyl invariant. It is the Liouville action $S_{\phi}$ that acts
as a Wess-Zumino-Witten term to recover the Weyl symmetry.}.

This counterterm can be removed, if one wishes, as follows.
By using the equations of motion in the conformal gauge, we first express
the counteraction
in terms of the phase space variable:
\begin{eqnarray}
S_{g}&=&-\frac{\kappa}{4} \int d^2\sigma \,
( V_{N}^{+} G_{-} +  V_{N}^{-} G_{+}
 + N^{+\prime} G_{+} + N^{-\prime} G_{-}) \ ,
\end{eqnarray}
where we have used the relations
\begin{eqnarray}
\sqrt{-g} R &=& -\partial_{\sigma}( V_{N}^{+} + V_{N}^{-} )
+ \frac{1}{2} \partial_{\tau}(G_{+} - G_{-})~, \nonumber\\
 V_{N}^{\pm} &=&
\frac{1}{2}\ G_{\pm} N^{\pm} + \partial_{\sigma} N^{\pm} ~,\nonumber\\
 G_{\pm} &=&  \frac{1}{{N}^{0}}~[~\pm  N^{\xi} + ({N}^{0} \mp
{N}^{1})~
\xi^{\prime} \mp 2~ {N}^{1 \prime}~]\ .
\end{eqnarray}
Further, the right-hand side of Eq.\ (49) can be written as
\begin{eqnarray}
\frac{i\kappa}{2}\,\int d^2\sigma [\eta~,~\Psi_{\rm CG}]\ ,
\end{eqnarray}
where $\Psi_{\rm CG}$ is the gauge fermion in the conformal gauge
($\Psi$ (23) with
$\chi$'s given in (31) and (38) ), and
\begin{eqnarray}
\eta &=& \frac{1}{4}\ G_{+}^2 {\cal C}^{+} + G_{+} ( {\cal C}^{+} )'
+\frac{1}{4}\ G_{-}^2 {\cal C}^{-} + G_{-} ( {\cal C}^{-} )'
 - \frac{1}{2} (G_{+} - G_{-}) {\cal C}^{\xi}\ .
\end{eqnarray}
Therefore, the counterterm $S_{g}$ can be canceled
by replacing the BRST charge
$\tilde{Q}$ by $\tilde{Q} - \kappa \int d \sigma~(\eta/2) $.
In configuration space, the redefined BRST charge is found to be
\begin{eqnarray}
\tilde{Q}_{\rm CG} &=& \tilde{Q} - \frac{\kappa}{2}
\int d \sigma~\eta  \nonumber\\
&=& \int d \sigma ~\sqrt{-\hat{g}}~\{~
[~\frac{C^0}{2} \hat{g}^{\alpha\beta}~
{\partial}_{\alpha} X^{\mu} {\partial}_{\beta} X_{\mu}
-{\hat{g}}^{0 \alpha} {\partial}_{\alpha}X^{\mu}~
C^{\gamma}{\partial}_{\gamma}X_{\mu}~] \nonumber\\
& & +\frac{\kappa}{2} ~[\frac{C^0}{2} {\hat{g}}^{\alpha\beta}~
{\partial}_{\alpha} \phi {\partial}_{\beta} \phi
-{\hat{g}}^{0\alpha}
{\partial}_{\alpha}\phi~C^{\gamma} {\partial}_{\gamma}\phi~]
\nonumber\\
& &+\frac{\kappa}{2}~[~
\frac{K^0}{\sqrt{-\hat{g}}}~C^{\gamma} {\partial}_{\gamma} \phi -
{\hat{g}}^{0\alpha} {\partial}_{\alpha}\phi~(
C^0\frac{{\hat{g}}_{01}}{\hat{g}_{11}} +C^1)~
\frac{\hat{g}_{11}^{\prime}}{\hat{g}_{11}}
\nonumber\\
& &
-2\,\hat{g}^{0\alpha}\partial_{\alpha}\phi~(
C^0\frac{{\hat{g}}_{01}}{\hat{g}_{11}}+C^1)^{\prime} +
 \frac{C^{0}}{\hat{g}_{11}}
-2{\phi}^{{\prime} {\prime}})~]
\nonumber\\
& &+(\overline{b}_{00} \hat{g}^{00}+
\overline{b}_{11} \hat{g}^{01})(C^{0}~{\dot{C}}^{1} + C^{1}~{C}^{1 {\prime}}) -
 \frac{\overline{b}_{11}}{\hat{g}_{11}}\,C^0
 {\dot{C}}^{0} \nonumber\\
& & + (\overline{b}_{01} \hat{g}^{01} -
\overline{b}_{11} \hat{g}^{11}) C^{1}~{C}^{0 {\prime}} ~\}~,
\end{eqnarray}
where
\begin{eqnarray}
& &\sqrt{-\hat{g}} \,R(\hat{g})~=
{}~\partial_{\alpha}K^{\alpha}~,\nonumber\\
K^0(\hat{g})
&=&\frac{1}{\sqrt{-\hat{g}}}\,(\,\dot{\hat{g}}_{11}-2\hat{g}_{01} '+
\frac{\hat{g}_{01}\hat{g}_{11} '}{\hat{g}_{11}}\,)~,~ K^1(\hat{g})\,=\,
\frac{1}{\sqrt{-\hat{g}}}\,(\,\hat{g}_{00} '-
\frac{\hat{g}_{01}\dot{\hat{g}}_{11}}{\hat{g}_{11}}\,)\ .\nonumber
\end{eqnarray}
To obtain the second equation in Eq. (53), we have
frequently used the equations of motion in the conformal gauge.
(The covariant quantities are defined in
section 4.) On the flat Minkowski space, the BRST current given in
Eq. (53)
reduces to the well
 known form
\begin{eqnarray}
 \tilde{J} &=& \{ ~C^{+}[~({\partial}_{+} X)^2 + \frac{\kappa}{2}
 ({\partial}_{+} \phi)^2 - \kappa~{\partial}_{+}^2 \phi +
\frac{{\mu}^2}{2} {\em e}^{\alpha\phi}
+\overline{b}_{++} {\partial}_{+} C^{+}]~\} \nonumber\\
&  &+ \{[+] \rightarrow [-]~\}~.
\end{eqnarray}

The re-definition of the BRST charge discussed above has the following
interpretation. If we remove all the BF fields,
the re-defined BRST charge develops a BRST anomaly,
which is not of the form given in Eq.  (15),  but it is
\begin{equation}
\Omega - K \int d \sigma ~\delta \eta~ ,
\end{equation}
where $\eta$ is defined in Eq. (52), and
the absence of the BF fields reduces the anomaly
coefficient from $\kappa$ to $K$.
  As shown in Ref.~\cite{fikm}, this shift by a coboundary
 term exactly gives rise to
the covariant
expressions for the BRST anomalies
\begin{eqnarray}
{\Omega}_{\rm{cov}} &=& K \int d \sigma
 ~ [~ \sqrt{-g} R C^0  C_{W} +  \sqrt{-g} g^{0 \alpha}
 C_{W} {\partial}_{\alpha}  C_{W}~ ]~, \nonumber\\
{\Gamma}_{\rm{cov}} &=& -K \int d \sigma
 ~\sqrt{-g} R C_{W} ~,
\end{eqnarray}
instead of the non-covariant expressions (15) and (24).
This means that, if the underlying regularization schemes
 respect the reparametrization invariance,
one may begin with the BRST anomalies given
by Eq. (56), and  end up with the covariant effective action
$S_{X}~ +~ S_{\phi}~ + ~ S_{gf+gh}$.

In summary, we have obtained the effective action in the conformal gauge,
$S_{X} + S_{\phi} +  S_{gf+gh}$, which is equivalent to the
the DDK action \cite{d,dk}, though these two actions
have a slightly different
cosmological constant term.
It should be remarked, however,
that in Ref.\ \cite{dk}
a conjecture was needed to use the translation
invariant measure for the Liouville mode
which is embedded as a component of the 2D
metric variables.
 In contrast to this, the Liouville mode $\phi$ in our approach
originates from the anomaly-compensating
degree of freedom, and
the functional measure in the path-integral quantization on the EPS
is fixed as the canonical measure
which is
translation invariant by construction.

\section{LIGHT-CONE GAUGE}
The light-cone gauge, $g_{+-} = -1/2,~g_{--} = 0$, is realized by
the gauge-fixing functions
\begin{eqnarray}
\chi^{+}&=&N^{+}-1~, \quad
\chi^{-} = [(N^{-}+1)e^\xi]~ - 2~,\nonumber\\
\chi^{\xi}&=&\xi-\theta~,
\end{eqnarray}
along with those given in Eq.\ (31).  The last condition in
Eq.\ (57) eliminates the  $\phi\,(=\xi-\theta)$
field, and instead, $g_{11} =
\exp\xi = g_{++} + 1$  behaves as the light-cone gauge analog
 of the Liouville mode.
As a result, the canonical structure becomes rather
 complicated, and we have to make a suitable change
 of the BFV variables to make our analysis simple.
  We shall begin by considering the classical theory, and establish the
$SL(2,R)$ Kac-Moody symmetry from the original BRST invariance
 already formulated on the BFV basis.
We then quantize the system
in the operator formalism to see whether we can
consistently maintain the current algebra.

To obtain the gauge-fixed action,
 we substitute the gauge fermion
(23) with $\chi$'s given in Eqs.\ (31) and (57) into the master action
(30), and then eliminate again all the non-dynamical fields
by using the equations of motion
such as
\begin{eqnarray}
\pi_{\theta} &=& \frac{\kappa}{2}~(~\partial_{-} g_{11} -
(~ \ln g_{11}~)^{\prime}~)~
 -~g_{11} \,\overline{{\cal C}}_{-}~
(~{\cal C}^{+} + {\cal C}^{-}~)~,
\nonumber\\
\overline{{\cal P}}_{+} &=&~
-\overline{{\cal C}}_{+}~,~\overline{{\cal P}}_{-}~=~-g_{11}
\overline{{\cal C}}_{-}~, \\
{\cal C}^{\xi} &=& \frac{1}{2}\, (~{\cal C}^{+} + {\cal C}^{-}~)
\partial_{-} g_{11} -  {\cal C}^{-} (~ \ln g_{11}~)^{\prime} +
 (~{\cal C}^{+} - {\cal C}^{-}~)^{\prime}~. \nonumber
\end{eqnarray}

The gauge-fixed action is then given by
\begin{eqnarray}
 S_{\rm LG}  & = & S_{X} + S_{g} + S_{gh}~, \nonumber\\
 S_{X} & = &
\int d^2 \sigma ~ \frac{1}{2}~[~
                  (~ g_{11}-1)~~ ({\partial_{-}X})^2
                   + \partial_{-} X \cdot \partial_{+}X ~)~]~,
\nonumber\\
 S_{g} & = & \int d^2 \sigma ~ \frac{\kappa}{4 g_{11}}~~
                  [~ (~ \partial_{-}g_{11})^2
               -  2~(~ \partial_{-}g_{11}~)(~ \ln g_{11} ~)^{\prime}
              +  4~(~ \ln g_{11}~)^{\prime \prime}  ~]~,
 \nonumber \\
S_{gh} & = & -\,\int d^2 \sigma~
[~b_{++} \partial_{-} c^{+} + b \partial_{-} c_{+}~]~,
\end{eqnarray}
where we have introduced
the new ghost variables
\begin{eqnarray}
c^{+} &=& {\cal C}^{+} ~,~
c_{+} ~=~{\cal C}^{-} + g_{++}~({\cal C}^{+} + {\cal C}^{-}) +
 \frac{\sigma^{-}}{2}\partial_{+}{\cal C}^{+}~,\nonumber\\
b_{++} &=& (\, \overline{{\cal C}}_+ -
g_{11}\overline{{\cal C}}_{-} +\frac{\sigma^{-}}{2}\partial_+
\overline{{\cal C}}_ +\,)~,~b~=~\overline{{\cal C}}_- ~,
\end{eqnarray}
in order to simplify the ghost sector.
($g_{++}=g_{11}-1=\exp\xi -1$)
The cosmological constant term becomes a constant
in the light-cone gauge (as shown in Appendix A), and so we
have suppressed it in the action.
The light-cone gauge action (59) is local, and should be compared
with the non-local action of Polyakov \cite{pol1}.
The form of $S_{X}$ and $S_{gh}$ may be expected from
a general consideration on the light-cone gauge, whatever
the $Q^2$-anomaly in Eq. (15) looks like. It is $S_{g}$
which contains the information of the anomaly, and
has not been derived before.

The equations of motion which follow from the action read:
\begin{eqnarray}
\partial_{-}\partial_{+}X^{\mu}&=&
-\partial_{-}(\, (g_{11}-1)\partial_{-}X^{\mu}\, )~,\nonumber\\
\partial_{-}\,c^+ &=& \partial_{-}\,c_{+}~=0~,~
\partial_{-}\,b_{++} ~=~ \partial_{-}\,b~=~0~, \nonumber\\
\frac{\kappa}{4}~\partial_{-}^{2}\, g_{11}
 & = &   \frac{g_{11}}{4}\, (\partial_{-}X)^2
    +\frac{\kappa}{2g_{11}}\,
        [~\frac{({\Theta^0_-})^2}{4}~-~({\Theta^0_-})^{~ \prime} ~]~ ,
\end{eqnarray}
where $
{\Theta^0_-} \equiv  2~(~\ln g_{11})^{\prime}~-~
\partial_{-}g_{11}~$.
Note that
the new ghosts satisfy the free equations of motion.
Rewriting (27) in terms of the new variables
and using the equations of motion (61),
we obtain the BRST transformations in the light-cone gauge
\begin{eqnarray}
\delta X &=& \frac{1}{2} c^{+} {\partial}_{+} X +
\frac{1}{2} (c_{+} - \frac{\sigma^{-}}{2}~{\partial}_{+} c^{+})~
{\partial}_{-} X~, \nonumber\\
\delta g_{++} &=& \frac{1}{2} c^{+} {\partial}_{+} g_{++} +
{\partial}_{+} c^{+} g_{++} + \frac{1}{2} (c_{+} -
\frac{\sigma^{-}}{2}~
{\partial}_{+} c^{+})~
{\partial}_{-} g_{++} - \frac{1}{2}{\partial}_{+} (c_{+} -
\frac{\sigma^{-}}{2}~{\partial}_{+} c^{+})~,\nonumber\\
\delta c^{+} &=& \frac{1}{2} c^{+} {\partial}_{+} c^{+}~, \\
\delta c_{+} &=& \frac{1}{2} c^{+} {\partial}_{+} c_{+} +
\frac{1}{2}{\partial}_{+} c^{+} c_{+}~, \nonumber\\
\delta b_{++} &=& T^{\rm total}_{++} \equiv T_{X} + T_{g} +
T_{b++} + T_{b}~, \nonumber\\
\delta b &=& \frac{\kappa}{4}\, {\partial}^2_{-} g_{++} +
\frac{{\mu}^2}{2} -
\frac{1}{2}\,
 {\partial}_{+} b~c^{+}~, \nonumber
\end{eqnarray}
where we have defined:
\begin{eqnarray}
T_{X} &\equiv& \frac{1}{4}~({\partial}_{+}X + g_{++}
{\partial}_{-}X)^2~,
\nonumber\\
T_{g} &\equiv& \frac{\kappa}{8}\,[\,({\partial}_{-} g_{++})^2 -
    2 \,g_{++}\,{\partial}^2_{-}g_{++} - (2 - \sigma^-
{\partial}_{-})\,
 {\partial}_{+}{\partial}_{-}g_{++}\,]~,
\\
T_{b++} &\equiv& - \frac{1}{2}\,{\partial}_{+} b_{++}\, c^{+}
- b_{++}\, {\partial}_{+} c^{+}~,
\nonumber\\
T_{b} &\equiv& \frac{1}{2}\,{\partial}_{+} b\, c_{+}~.
\nonumber
\end{eqnarray}
In terms of these stress tensors, the BRST charge can be written as
\begin{eqnarray}
{\tilde Q}_{\rm LG} &=&
\int d \sigma^+~ [~c^{+}~(T_{X} + T_{g} + \frac{1}{2}
T_{b++} + T_{b}) +
 c_{+}~( \frac{\kappa}{4}\,{\partial}^2_{-} g_{++} +
\frac{{\mu}^2}{2})~]~.
\end{eqnarray}
This coincides with the BRST charge of Ref.\ \cite{kim} except for
the contribution from the cosmological constant term.

The $SL(2,R)$ Kac-Moody symmetry in the light-cone gauge
arises
in the present formalism as follows:
First, one derives Polyakov's fundamental identity,
  \cite{pol1,kpz}
\begin{eqnarray}
\partial_{-}^{3}~g_{++} &=& 0~,
\end{eqnarray}
 by considering the BRST variation of
the equation of motion, $\delta (\partial_{-}~b)~=~0~$.
Then one expands the gravitational field $g_{++}$ according to
\begin{eqnarray}
g_{++} &=& -~\frac{1}{2\kappa}~[~J^{+}(\sigma^{+})-2\sigma^{-}
J^{0}(\sigma^{+})+(\sigma^{-})^2 J^{-}(\sigma^{+})~]~.
\end{eqnarray}
Substituting Eq. (66) into $T_{g}$ given in Eq. (63), one finds that
 the stress tensor of the
 gravity sector takes the Sugawara form:
\begin{eqnarray}
T_{g} &=& \frac{1}{2\kappa} ~[~(J^{0})^2-J^{+}J^{-}~]-\frac{1}{2}\,
\partial_{+}J^0~.
\end{eqnarray}
At the same time, one also expands
 $\delta g_{++}$ in Eq. (62) in powers of $\sigma^-$ to find the BRST
 transformation of the currents:
\begin{eqnarray}
\delta J^{+} &=& \frac{1}{2}~c^{+} \partial_{+} J^{+}~+
{}~\partial_{+} c^{+}
 J^{+}~-~2~c_{+} J^{0}~+~\kappa\, \partial_{+} c_{+}~, \nonumber\\
\delta J^{0} &=& \frac{1}{2}~c^{+} \partial_{+} J^{0}~+
 \partial_{+} c^{+}
 J^{0}~-~c_{+} J^{-}~+~\frac{\kappa}{4}\, {\partial}^2_{+} c^{+}~, \\
\delta J^{-} &=& \frac{1}{2}~c^{+} \partial_{+} J^{-}~.  \nonumber
\end{eqnarray}
In Appendix B it is shown how these $\delta\,J^a$ imply
 the $SL(2,R)$ current algebra
\begin{eqnarray}
\{J^{a}(\sigma^+)~,~J^{b}(\sigma^{\prime+})\}_{\rm PB} &=&
f^{ab}{}_{c}J^{c}(\sigma^+)
\delta (\sigma^{+} - \sigma^{\prime+}) - \kappa\,
\eta^{ab}\, \delta^\prime (\sigma^{+} - \sigma^{\prime+})~,
\end{eqnarray}
where $f^{ab}{}_{c} = - f^{ba}{}_{c}$
 and
$\eta^{ab} =\eta^{ba}$ have non-vanishing
 components,
$f^{+0}{}_{+} = f^{0-}{}_{-}=1~,~f^{+-}{}_{0} =2~,~\eta^{00}=-1~,~
\eta^{+-}=2$.

Now to achieve the full quantum treatment of the system, we first
assume
 that the current algebra (69) is realized under commutator too:
\begin{eqnarray}
[~J^{a}(\sigma^{+})~,~J^{b}(\sigma^{\prime+})~] &=& i~
f^{ab}{}_{c}~J^{c}(\sigma^{+})~
\delta (\sigma^{+} - \sigma^{\prime+})~ -~ i~\kappa~ \eta^{ab}~
\delta^\prime (\sigma^{+} - \sigma^{\prime+})~.
\end{eqnarray}
And to find the quantum operator corresponding to $T_{g}$, we
consider \begin{eqnarray}
 T & \equiv & \eta_{ab} :J^{a}~J^{b}: = \eta_{ab}
\,(J^{a,-}J^{b,-}~+~J^{a,-}J^{b,+}~+~ J^{a,+}J^{b,+}~+~
J^{b,-}J^{a,+}~)~,
\end{eqnarray}
where the normal ordering prescription is used to regularize the
 products of the currents (see Appendix C).
The algebraic relations among $J^{a}$ and $T$ given in
Eq. (A.10) in Appendix C
 imply that the quantum operator for $T_{g}$ has to be defined as
\begin{eqnarray}
T_{g}(\sigma^{+}) &=& - \frac{\pi}{2\pi\kappa - 1}~T(\sigma^{+})~-
\frac{1}{2}~\partial_{+} J^{0}(\sigma^{+})~.
\end{eqnarray}
Then from the Sugawara construction, one obtains the commutation relations
\begin{eqnarray}
[J^{a}(\sigma^{+})~,~T_{g}(\sigma^{\prime+})] &=&
\frac{i}{2}\,\partial_{+}\{ J^{a}(\sigma^{+})\,
\delta (\sigma^{+}-\sigma^{\prime+})\}+~
i f^{a0}{}_{b}\, J^{b}(\sigma^{+}) \,
\delta^\prime (\sigma^{+}-\sigma^{\prime+}) \nonumber\\
& &~-~i~\kappa~\eta^{a0} \,
\delta^{\prime\prime} (\sigma^{+}-\sigma^{\prime+})~,
\end{eqnarray}
\begin{eqnarray}
[~T_{g}(\sigma^{+})~,~T_{g}(\sigma^{\prime+})~] &=&
i\left(\, T_{g}(\sigma^{+}) + T_{g}(\sigma^{\prime+}) \,\right)
\delta^\prime (\sigma^{+}-\sigma^{\prime+})\nonumber\\
& & - i\, \frac{c_{g}}{24\pi}~
\delta^{\prime\prime\prime} (\sigma^{+}-\sigma^{\prime+})\ .
\end{eqnarray}
The central charge $c_{g}$ is given by
\begin{eqnarray}
c_{g} &=& \frac{6\pi\kappa}{2\pi\kappa-1}~+~24\pi\kappa
 = \frac{3k}{k-2}~+~6k~,
\end{eqnarray}
where $k \equiv 4\pi\kappa $ is the central charge of the current
 algebra.
 The stress tensors for the matter and  ghost sectors contained
in the BRST charge operator
\begin{eqnarray}
{\tilde Q}_{\rm LG} &=& \int d \sigma^+~
[~c^{+}~(T_{X} + T_{g} + \frac{1}{2} T_{b++}
 + T_{b}) -  c_{+}~( J^{-} - \frac{{\mu}^2}{2})~]
\end{eqnarray}
have the central charges,
$c_{X} = D$ and $c_{gh} = -28$, respectively.
Therefore, the quantum nilpotency condition,
$[{\tilde Q}_{\rm LG}~,~{\tilde Q}_{\rm LG}] = 0$ implies
 the KPZ condition \cite{kpz}
\begin{eqnarray}
\frac{3k}{k-2}~+~6k~+~D~-28 &=& 0~.
\end{eqnarray}

Before we close this section, we would like to summarize the content
of this section. We have derived the local gauge-fixed action (59)
from the master action (30). Then using the conservation
of the BRST charge (26) and the equation of motion for the $b$ ghost
(see Eq. (61)), we have obtained the constant curvature equation (65),
which makes the decomposition of $g_{++}$ into the chiral currents
$J^a$ possible. Since $g_{++}=g_{11}-1=\exp \xi -1$,
the BRST transform of $g_{++}$ is already known. Expressing
$\delta\,g_{++}$ in terms of $J^a$, we have deduced $\delta J^a$,
which is given in Eq. (68). At the same time, we have re-written
the gauge-fixed form of the BRST charge (64) in terms of $J^a$,
and compared
$\{J^a~,~\tilde{Q}_{\rm LG}\}_{\rm PB}$ with the
$\delta\,J^a$ which has been deduced from $\delta\,g_{++}$. In doing
so, we have derived the $SL(2,R)$ current algebra (69)
at the level of Poisson bracket.
We may fairly say that the $SL(2,R)$ Kac-Moody symmetry is
 to be traced back
to the original BRST symmetry, which is defined
on the extended phase space before the gauge-fixing.

Having arrived at this stage, one can follow Refs. \cite{pol1,kpz,kim}
for the rest of the analyses. As a by-product, we have found that
a cosmological constant can be included in the light-cone
gauge and it does not affect the KPZ condition (77)
(if the relation between $\alpha$ and $\kappa$ given in Eq. (44)
is satisfied in the light-cone gauge too).

\section{HARMONIC GAUGE}
In the BFV formalism it is also possible to take the
harmonic gauge \cite{an}
where
 the metric variables become dynamically active.  To realize such gauge,
we must keep the Legendre terms for the multiplier fields in the
 master action.
Let us first consider the gauge-fixing corresponding to
 reparametrization symmetry.
The relevant part in the master action is given by
\begin{eqnarray}
S_{gf+gh}^R &=& \int d^2 \sigma ~(~
{\cal B}_{\pm} {\dot{N}}^{\pm} + \overline{{\cal C}}_{\pm}
\dot{\cal{P}}^{\pm}
- {\cal B}_{\pm} {\chi}^{\pm} - \overline{{\cal C}}_{\pm}
\delta {\chi}^{\pm} ~)~,
\end{eqnarray}
where the summation is taken over the constraint-label $\pm$.
We would like this action to be
written in terms
of the GL(2)-covariant variables, and to identify it
 with the gauge-fixing action \cite{an}
\begin{eqnarray}
 \int d^{2}\sigma ~[~ B_{\alpha} {\partial}_{\beta}
(\sqrt{-g} g^{\alpha \beta}) +
\overline{C}_{\alpha} \,{\partial}_{\beta} \delta
(\sqrt{-g} g^{\alpha \beta})~]~,
\end{eqnarray}
where the BRST transformations for the covariant variables
are given in Eqs. (36) and (37).
The action (78) indeed
becomes identical to (79), if we require the relations
\begin{eqnarray}
B_{0} &=& N^{0}(N^{0}~{\cal B}_{0} + N^{1}~{\cal B}_{1}) +
2\,(N^{0}~ \overline{{\cal C}}_{0} + N^{1}~
\overline{{\cal C}}_{1})~{\cal P}^{0}\nonumber\\
 & &+ \overline{{\cal C}}_{1}\,(N^{0}~{\cal P}_{1} -
N^{1}~{\cal P}_{0})~,
\nonumber\\
B_{1} &=& N^{0}~{\cal B}_{1} + \overline{{\cal C}}_{1} {\cal P}_{0}~,
\nonumber\\
\overline{C}_{0} &=& N^{0}\,(N^{0}~ \overline{{\cal C}}_{0} + N^{1}~
\overline{{\cal C}}_{1})~,
\nonumber\\
\overline{C}_{1} &=& N^{0}~\overline{{\cal C}}_{1}~,
\end{eqnarray}
and the gauge-fixing functions are given by
\begin{eqnarray}
{\chi}^{0} &=& N^{1}~N^{0 {\prime}} - N^{0}~N^{1 {\prime}}~, \quad
{\chi}^{1} = N^{1}~N^{1 {\prime}} - N^{0}~N^{0 {\prime}}~.
\end{eqnarray}

As for the gauge-fixing corresponding to
the Weyl symmetry, we may impose the flat-curvature condition given
by
\begin{eqnarray}
S_{gf+gh}^W &=& \int d^{2}\sigma ~[\, B_{W} \sqrt{-g}~R +
\overline{C}_{W} \delta (\sqrt{-g}~R)\,]~.
\end{eqnarray}
To see that the covariant action (82) can be
obtained from the one defined in terms of the non-covariant BFV variables,
we use the non-covariant decomposition of the curvature:
\begin{eqnarray}
\sqrt{-g}~R &=& {\partial}_{\alpha} K^{\alpha}~,
\nonumber\\
K^{0} &=& \frac{1}{N^{0}}~{\dot{N}}^{\xi} + L~,
\nonumber\\
K^{1} &=& -\frac{1}{N^{0}}[\,N^{1} N^{\xi} + N^{+} N^{-} {\xi}^{\prime} +
(N^{+} N^{-})^{\prime}\,]~,
\end{eqnarray}
 where $N_{\lambda}^{0,1} = {\dot{\lambda}}^{0,1} =\dot{N}^{0,1}$, and
\begin{eqnarray}
L &= &\frac{1}{{N^{0}}^2}~[\,(N^{0}-N_{\lambda}^{0})\, N^{\xi}
+(N^{1} N_{\lambda}^{0} -N^{0} N_{\lambda}^{1})\,{\xi}^{\prime}
- N^{0} N^{1}
N^{\xi {\prime}} \nonumber\\
 & & \quad +2 (\,N^{1 {\prime}} N_{\lambda}^{0} -N^{0} N_{\lambda}^{1
{\prime}}\,)\,]~.
\end{eqnarray}
Imposing the relations
\begin{eqnarray}
B_{W} &=& {\cal B}_{\xi}\, N^{0} +
\overline{{\cal C}}_{\xi}\, {\cal P}^{0},
 \quad \overline{C}_{W} = N^{0} \,\overline{{\cal C}}_{\xi}~,\nonumber\\
{\chi}^{\xi} &=& - N^{0}(L + K^{1 '})~,
\end{eqnarray}
one finds that
the BFV action
\begin{eqnarray}
S_{gf+gh}^W &=& \int d^{2}\sigma ~(\,
{\cal B}_{\xi} {\dot{N}}^{\xi} + \overline{{\cal C}}_{\xi}
\dot{{\cal P}}^{\xi}
- {\cal B}_{\xi} {\chi}^{\xi} - \overline{{\cal C}}_{\xi}
\delta {\chi}^{\xi}\, )
\end{eqnarray}
is identical to the covariant one (82).

The total action is then given by $S_{\rm HG} =
S_{X} + S_{\phi} + S_{g} + S_{gf+gh}^R +
S_{gf+gh}^W$.
  We will not discuss here the full quantum treatment of
 the theory in the harmonic gauge, though it is
certainly an interesting problem.

\section{DISCUSSIONS AND SUMMARY}
We have treated 2D gravity as an anomalous
gauge theory and applied the canonical formalism of Refs.\ \cite{mo,fik1}
which is based on the extended
phase space method of BF \cite{bf}. In doing so, we are able to
formulate 2D gravity in most general class of gauges.
The conformal, light-cone, and
harmonic gauges have been explicitly considered
to illustrate how to get the gauge-fixed action in those gauges
from the master action (30). This derivation of the gauge-fixed
theories in 2D gravity might be related to
the observation of Ref.\ \cite{fw} that the
Wess-Zumino-Witten model based on $SL(2,R)$
plays a crucial r{\^ o}le to understand the relation
between those theories.

It is well known that in the path-integral quantization
on configuration space, one encounters the ambiguity in defining a
local measure.  A possible way to
 avoid it is to impose some invariance on the measure.
For 2D gravity, the measure is usually fixed so as to be
 reparametrization invariant.  Without the anomaly-compensating fields,
 it would definitely
lead to a complicated, translation non-invariant measure for the
conformal factor of the metric.
This is the origin of the DDK conjecture.
 The path-integral quantization based on the EPS, however,
does not suffer from this problem, because
that measure in this approach is uniquely
fixed as the Liouville measure which is translation invariant
by construction.

The presence of the non-covariant counterterm $S_{g}$ in the
conformal gauge
indicates that the path-integral measure to be used for this case
is not reparametrization invariant.
The path-integral Jacobian associated with the transition to
the reparametrization invariant measure can be calculated by
using Fujikawa's method \cite{fuji} in principle.
At least in the $\hat{N}^{\pm}=1$
gauge, we can verify that the Jacobian in question
exactly cancels $S_{g}$.
 This is another derivation of the
equivalence between the DDK approach and ours. The derivation
based on the operator language has been presented in section 5.

We have delivered
a BRST formulation of 2D gravity in the light-cone gauge,
thereby clarifying the origin of the $SL(2,R)$ current algebra.
It is nothing but
the original BRST invariance from which the $SL(2,R)$
Kac-Moody symmetry (69) originates,
as one can convince oneself from our derivation
of the $SL(2,R)$ current algebra in section 6.
It should be reminded once again that,
in the previous derivations of the BRST charge \cite{kim},
one either starts with Polyakov's non-local, light-cone effective
action which has an $SL(2,R)$ symmetry,
or bases on an $SL(2,R)$ current algebra from the beginning.

We have shown that the $SL(2,R)$ symmetry can be maintained in the full
quantum treatment in accord with the result of Refs. \cite{pol1,kpz,kim}.
We also have shown,
both in the conformal and light-cone gauges, that the quantum nilpotency
condition on the
 BRST charge is not affected in the presence of a cosmological
constant term.

 In conclusion, our approach
to 2D gravity gives a common basis to formulate the theory in different
gauges, and is therefore suitable to study different
dynamical aspects of the theory;
it is certainly useful to analyze 2D cosmology
\cite{cos} in different gauges for
instance. An extension of our approach to 2D chiral gravity
\cite{cg} is straightforward.

\vspace{5mm}

\noindent
{\bf ACKNOWLEDGMENT}
\\
We would like to thank K. Fujikawa, K. Itoh and
H. Terao for useful discussions
and comments.

\newpage
\begin{center}{\bf APPENDIX}
\end{center}
Here we derive the algebraic identities in the light-cone gauge,
that have been used in section 6.
\newline
\vspace{5mm}

\noindent{\bf A.~The cosmological constant term} \\
Let us first show that the cosmological constant term becomes trivial
in the light-cone gauge.
  Consider the BRST variation of ${\cal V} = e^{\alpha \theta}$ and
$g_{11}= \exp(\xi)$.  We regularize ${\cal V}$ by taking the
 normal ordering
to obtain
$$
-i [~{\tilde Q}~,~:{\cal V}:~] =
 C^{\alpha}\partial_{\alpha}
 :{\cal V}: + \frac{2}{g_{11}} C^{\alpha \prime} g_{\alpha 1}:{\cal V}:~,
\eqno{({\rm A}.1)} $$
$$
\delta e^{\xi} = e^{\xi}~ {\cal C}^{\xi} = C^{\alpha}\partial_{\alpha}
 g_{11} + 2 C^{\alpha \prime} g_{\alpha 1}~,
\eqno{({\rm A}.2)} $$
where we have used the relation (44),
the equations of motion in the light-cone
gauge (58), and the relations for the
ghosts  in that gauge
$$
{\cal C}^{+} = C^{0}~+~C^{1}~,\quad
{\cal C}^{-} = (\, \frac{2}{g_{11}}-1\,) C^{0}-C^{1}.
\eqno{({\rm A}.3)} $$
(See Eqs. (33) and (35) )
{ }From Eqs. ({\rm A}.1) and ({\rm A}.2) one sees that
${\cal V}$ and $g_{11}= \exp(\xi)$ have the same transformation
property, and
therefore,
one should identify ${\cal V}$ with $g_{11}= \exp(\xi)$
though the gauge condition in the light-cone gauge
$ \theta =\xi$.
  It follows then that the cosmological constant
term becomes really constant because
$$
S_{\rm cosm} = - \frac{\mu^2}{2} \int d^2 \sigma (~N^{-} +1~) e^{\alpha
\theta} = - {\mu}^2 \int d^2 \sigma \frac{e^{\alpha \theta}}{g_{11}}~,
\eqno{({\rm A}.4)} $$
and does not contribute to the the equation of motion for $g_{11}$.
It does, however, contribute to the BRST charge as given in
Eq. (64):
$$
\tilde {Q}_{\rm cosm} =  \frac{\mu^2}{2} \int d\sigma (~{\cal C}^{+} +
 {\cal C}^{-}~)\, e^{\alpha \theta} $$ $$
= \frac{\mu^2}{2} \int d\sigma (~c_{+} + c^{+} - \frac{\sigma^{-}}{2}
\partial_{+} c^{+}~)
= \frac{\mu^2}{2} \int d \sigma^+\, c_{+}~,
\eqno{({\rm A}.5)}$$
where we have used Eqs. ({\rm A}.3), (60), and the equations of motion
 (61).

\newpage
\noindent
{\bf B. The derivation of the $SL(2,R)$ current algebra} \\
The relevant part of the BRST charge $\tilde{Q}_{\rm LG}$
 can be written in terms of the currents $J^a$:
$$
{\tilde Q}_{g} = - \int d \sigma^+ [\, c^{+}(~ \frac{1}{2\kappa}
\eta_{ab} J^{a}J^{b} +  \frac{1}{2} \partial_{+} J^{0}~)
+c_{+}\,J^-\,]~.\eqno{({\rm A}.6)}
$$
The BRST transform of $J^a$, which has been read off
from $\delta\,g_{++}$ in Eq. (62), is given in Eq. (68), on one hand.
On the other hand, $\delta\,J^a$ is defined by
 $$
\delta J^{a}(\sigma^+) =
\{~J^{a}(\sigma^+)~,~{\tilde Q}_{g}~\}_{\rm PB} $$ $$
= - \int d \sigma^{+ '}\,
[~ \frac{1}{\kappa} c^{+}(\sigma^{\prime+})~\eta_{bc}\,
\{J^{a}(\sigma^{+})~,~J^{b}(\sigma^{\prime+})
\}_{\rm PB}~J^{c}(\sigma^{\prime+}) $$
$$ +~c^{+}(\sigma^{\prime+})\,
\{J^{a}(\sigma^{+})~,~\frac{1}{2}\partial_{+}
J^{0}(\sigma^{\prime+})+ J^{-}(\sigma^{\prime+}) \}_{\rm PB}~]~.
\eqno{({\rm A}.7)} $$
Comparing this with $\delta J^{a}$ given in
Eq. (68) and assuming
 that the brackets of  the currents become at most linear in $J^{a}$, we
obtain the Poisson bracket relation (70).
\newline
\vspace{5mm}

\noindent
{\bf C. The Sugawara construction for the quantum $SL(2,R)$ current
algebra}
\\
We define the positive and negative frequency parts of the currents
by
$$
J^{a,\pm} (\sigma^{+}) = \int d\sigma^{\prime+}\,
{\delta}^{(\mp)}(\sigma^{+}-\sigma^{\prime+})\,
J^{a}(\sigma^{\prime+})~,  \eqno{({\rm A}.8)}$$ $$
{\rm with}~~ {\delta}^{(\pm)}(\sigma^{+}) =
 \frac{\pm i}{2\pi(\sigma^{+} \pm i0+)}~,
$$
and employ the normal ordering prescription
(with respect to $J^{a,\pm}$) to define the products
of the currents.
The assumption that the current algebra (69) persists at the quantum level
yields
$$
[~J^{a}(\sigma^{+})~,~J^{b,\pm}(\sigma^{\prime+})~] = i f^{ab}{}_{c}\,
J^{c}(\sigma^{+})\, \delta^{(\pm)} (\sigma^{+}-\sigma^{\prime+})~-~i \kappa\,
\eta^{ab}\, \delta^{(\pm)\prime} (\sigma^{+}-\sigma^{\prime+})~,
\eqno{({\rm A}.9)} $$
which can be used
to obtain the commutator with $T=\eta_{ab}:J^{a} J^{b}:$
$$
[~J^{a}(\sigma^{+})~,~T(\sigma^{\prime+})~] = -\,i~ \frac{2\pi\kappa
-1}{2\pi}\, \partial_{+}
\{ J^{a}(\sigma^{+})\, \delta (\sigma^{+}-\sigma^{\prime+})\}~.\eqno{({\rm
A}.10)} $$
If we define
$$
{\hat T} = -\frac{\pi}{2\pi\kappa-1}~T~,\eqno{({\rm A}.11)} $$
we find the commutation relation
$$
[{\hat T}(\sigma^{+})~,~{\hat T}(\sigma^{\prime+})]
= i \, (\,{\hat T}(\sigma^{+})+{\hat T}(\sigma^{\prime+})\,)
\delta^\prime (\sigma^{+}-\sigma^{\prime+})~
- \frac{i\kappa}{4(2\pi\kappa -1)}\,
 \delta^{\prime\prime\prime} (\sigma^{+}-\sigma^{\prime+})~.
\eqno{({\rm A}.12)} $$
These relations give rise to the fundamental commutators (73)
and (74) that are needed
 to show the nilpotency condition of the BRST charge
$\tilde{Q}_{\rm LG}$ (76).


\newpage

\begin{thebibliography}{99}
\bibitem{pol1} A. M. Polyakov, Mod. Phys. Lett. {\bf A2}, 893 (1987).

\bibitem{kpz}
V. G. Knizhnik, A. M. Polyakov, and A. B. Zamolodchikov, Mod. Phys. Lett.
{\bf A3}, 819 (1988).

\bibitem{d}F. David, Mod. Phys. Lett. {\bf A3}, 1651 (1988).

\bibitem{dk}J. Distler and H. Kawai, Nucl. Phys. {\bf B321}, (1989) 509.

\bibitem{fis}
K. Fujikawa, T. Inagaki and H. Suzuki, Phys. Lett. {\bf B213}, 279 (1988).

\bibitem{mmh}
N. Mavromatos and J. Miramontes, Mod. Phys. Lett, {\bf A4}, 1849
(1989);
\newline
E. D'Hoker and P. S. Kurzepa, Mod. Phys. Lett. {\bf A5}, 1411 (1990);
\newline
E. D'Hoker, Mod. Phys. Lett. {\bf A6}, 745 (1991).

\bibitem{mm}See for instance: Refs.\ \cite{dk,fis,mmh,fw}.

\bibitem{fw}
 A. Alekseev and S. Shatasvili,
Nucl. Phys. {\bf B323}, 719 (1989);
\newline
P. Forg{\' a}cs, A. Wipf, J. Balog, L. Feh{\' e}r, and L. O'Raifertaigh,
Phys. Lett. {\bf B227}, 214 (1989).


\bibitem{pol2}A. M. Polyakov, Phys. Lett. {\bf B103}, 207 (1981).

\bibitem{ct}T. L. Curtright and C. B. Thorn, Phys. Rev. Lett. {\bf 48},
1309 (1982);
\newline
 E. D'Hoker and R. Jackiw, Phys. Rev. {\bf D12},
3517 (1982);
\newline
D. S. Hwang, phys. Rev. {\bf D28}, 2614 (1983);
\newline
R. Marnelius, Nucl. Phys. {\bf B211}, 14 (1983);
\newline
J. -L. Gervais and A. Neveu, Phys. Lett. {\bf B151}, 271 (1985).

\bibitem{fikmt}T. Fujiwara, Y. Igarashi,
J. Kubo, K. Maeda and T. Tabai, Kanazawa University,
Report No. KANAZAWA 92-15, hep-th/9302130.

\bibitem{anti}I.A. Batalin and G.A. Vilkovisky,
 Nucl. Phys. {\bf B234} (1984) 106; Phys. Rev. {\bf D28}
(1983) 2567; {\bf D30} (1984) 508 (E)

\bibitem{gomis}
J. Gomis and J. Paris, Barcelona University, Report No. UB-ECM-PF-92-10,
hep-th/9204065.

\bibitem{fikm}T. Fujiwara, Y. Igarashi,
J. Kubo, and K. Maeda, hep-th/9210038, to appear in Nucl. Phys. B .

\bibitem{bfv}E. S. Fradkin and G. A. Vilkovisky, Phys. Lett. {\bf B55}, 224
(1975) ;
\newline
 I. A. Batalin and G. A. Vilkovisky, Phys. Lett. {\bf B69}, 309
(1977).

\bibitem{fs}
L.D. Faddeev and S.L. Shatashvili, Phys. Lett. {\bf B167}, 225 (1986).

\bibitem{mo}M. Moshe and Y. Oz, Phys. Lett. {\bf B224}, 145 (1989).


\bibitem{fik1}T. Fujiwara, Y. Igarashi and J. Kubo, Nucl. Phys. {\bf B341},
695 (1990); Phys. Lett. {\bf B251}, 427 (1990).

\bibitem{bf}I. A. Batalin and E. S. Fradkin, Phys. Lett. {\bf B180}, 157
(1986); Nucl. Phys. {\bf B279}, 514 (1987).

\bibitem{an} See, for example, M. Abe and N. Nakanishi, Int. J. Mod. Phys.
 {\bf A6}, 3955 (1991).

\bibitem{h}See also: M. Henneaux, Phys. Rep. {\bf 126}, 1 (1985).

\bibitem{gsw}M. B. Green, J. H. Schwarz, and E. Witten, {\em
Superstring Theory}, (Cambridge Univ. Press, 1987).

\bibitem{ko}M. Kato and K. Ogawa, Nucl. Phys.  {\bf B212}, 443 (1983).

\bibitem{fik}T. Fujiwara, Y. Igarashi and J. Kubo, Nucl. Phys. {\bf B358},
195 (1991).

\bibitem{fuji}K. Fujikawa, Phys. Rev. {\bf D25}, 2584 (1982).

\bibitem{kim}T. Kuramoto, Phys. Lett. {\bf B233}, 363 (1989);
\newline
K. Itoh, Nucl. Phys. {\bf B342}, 449 (1990);
\newline
Z. Horv\'{a}th, L. Palla and P. Vecserny\'{e}s,
Int. J. Mod. Phys. {\bf A4}, 5261 (1989);
\newline
N. Marcus and Y. Oz, Tel-Aviv University, Report No. TAUP-1962, 1992.

\bibitem{cos}
G. Mandal, A. Sengupta and S. Wadia, Mod. Phys. Lett.
{\bf A 6} (1991) 1685;
\newline
E. Witten, Phys. Rev. {\bf D44}, 314 (1991);
\newline
C. Callen, S. Giddings, J. A. Harvey, and A. Strominger, Phys. Rev.
{\bf D45}, 1005 (1992)

\bibitem{cg}Y. Oz, J. Pawelczyk, and S. Yankielowicz, Phys.
Lett. {\bf B249}, 417 (1990);
\newline
R. C. Myers and V. Periwal, McGill University, Report No. McGill/92-26.

\end{thebibliography}
\end{document}